\documentclass[preprint,12pt]{elsarticle}

\usepackage{amssymb}
\usepackage{amsthm}

\usepackage{color}

\journal{XXX}

\begin{document}

\begin{frontmatter}



\title{Exploiting dynamical perturbations for the end-of-life disposal of spacecraft in LEO}


\author[au1]{Giulia Schettino\corref{cor1}} 
\ead{g.schettino@ifac.cnr.it}
\author[au1]{Elisa Maria Alessi}
\ead{em.alessi@ifac.cnr.it}
\author[au1]{Alessandro Rossi} 
\ead{a.rossi@ifac.cnr.it}
\author[au1,au2]{Giovanni B. Valsecchi}
\ead{giovanni@iaps.inaf.it}

\cortext[cor1]{Corresponding author}
\address[au1]{IFAC-CNR, Via Madonna del Piano 10, 50019 Sesto Fiorentino (FI) - Italy}
\address[au2]{IAPS-INAF, Via Fosso dei Cavalieri 100, 00133 Rome - Italy}

\begin{abstract}
  As part of the dynamical analysis carried out within the Horizon
  2020 ReDSHIFT project, this work analyzes the possible strategies to
  guide low altitude satellites towards an atmospheric reentry through
  an impulsive maneuver. We consider a fine grid of initial conditions
  in semi-major axis, eccentricity and inclination and we identify the
  orbits that can be compliant with the 25-year rule as the target of
  a single-burn strategy. Besides the atmospheric drag, we look for
  the aid provided by other dynamical perturbations -- mainly solar
  radiation pressure -- to facilitate a reentry. Indeed, in the case
  of typical area-to-mass ratios for objects in LEO, we observed that
  dynamical resonances can be considered only in combination with the
  atmospheric drag and for a very limited set of initial
  orbits. Instead, if an area augmentation device, as a solar sail, is
  available on-board the spacecraft, we verified that a wider range of
  disposal solutions become available. This information is exploited
  to design an improved mitigation scheme, that can be applied to any
  satellite in LEO.
\end{abstract}


\end{frontmatter}


\section{Introduction}
\label{sec:intro}

The ``Revolutionary Design of Spacecraft through Holistic Integration
of Future Technologies'' (ReDSHIFT) project, funded by the H2020 Space
Work Program, addresses the topic of passive means to reduce the
impact of space debris. In the context of the compelling issue of
space debris mitigation, the project covers all the aspects of
planning a ``space debris tested'' mission, from a theoretical to
a technological and legal perspective \cite{RossiSD}.

In the following, we focus on the dynamical issues related to the Low
Earth Orbit (LEO) region. Recently, we performed an accurate mapping
of the LEO phase space, as described in \cite{AlessiSD, AlessiIAC,
  AlessiCMDA,Menios}, and we identified stable and unstable regions,
where dynamical perturbations as solar radiation pressure (SRP) and
lunisolar effects induce a relatively significant growth in
eccentricity, which can assist reentry. In this work, we complement
the analysis focusing on the possible reentry strategies for the
end-of-life disposal of spacecraft from LEO, by applying one impulsive
maneuver and, possibly, exploiting dynamical perturbations. The most
suitable maneuver is identified in terms of the minimum $\Delta v$
which ensures to be compliant with the well-known 25-year rule
\cite{IADC}.

The design of the transfer towards the Earth is based on the outcome
of the dynamical mapping already performed. The achieved results will
be used in the following in two ways: a first methodology considers,
as target orbits, the orbits which actually reenter in the desired
time span, among those explored in the cartography; a second
methodology takes advantage of the theoretical findings derived from
the cartography and aims at defining the most convenient reentry
trajectory following such information.

The paper is organized as follows: in Sec. \ref{sec:cart} we briefly
recall the basis of the cartography of the LEO phase space; in
Sec. \ref{sec:strat1} we describe the single-burn disposal strategy
based on the orbital grid defined for the cartography together with
some first results, while in Sec. \ref{sec:life} we analyze a
complementary strategy based on the theoretical explanation of the
results of the numerical mapping. Finally, in Sec. \ref{sec:concl} we
present a general discussion and draw some conclusions.

\section{Cartography of the LEO phase space}
\label{sec:cart}

As explained in \cite{AlessiSD, AlessiIAC, AlessiCMDA},
within the scope of ReDSHIFT we performed an extensive study of the
dynamics of the LEO region by propagating a fine grid of initial
orbits, selecting the initial semi-major axis $a$, eccentricity $e$
and inclination $i$ as shown in Table \ref{grid}.
\begin{table}[h!]
\begin{center}
\begin{tabular}{cccccc}
\hline
$a$ (km) & $\Delta a$ (km) & $e$ & $\Delta e$ & $i$ (deg) & $\Delta i$ (deg) \\
\hline
$R_E+[500-700]$ & 50 & $0-0.28$ & 0.01 & $0-120$ & 2 \\
$R_E+[720-1000]$ & 20 & $0-0.28$ & 0.01 & $0-120$ & 2 \\
$R_E+[1050-1300]$ & 50 & $0-0.28$ & 0.01 & $0-120$ & 2 \\
$R_E+[1320-1600]$ & 20 & $0-0.28$ & 0.01 & $0-120$ & 2 \\
$R_E+[1650-2000]$ & 50 & $0-0.28$ & 0.01 & $0-120$ & 2 \\
$R_E+[2100-3000]$ & 100 & $0-0.28$ & 0.01 & $0-120$ & 2 \\
\hline
\end{tabular}
\caption{Grid of the initial semi-major axis $a$, eccentricity $e$ and inclination $i$ adopted for the numerical propagation for the LEO mapping (see \cite{AlessiSD, AlessiIAC, AlessiCMDA}). $R_E$ refers to the Earth radius.} \label{grid}
\end{center}
\end{table}
Regarding the longitude of the ascending node $\Omega$ and argument of
pericenter $\omega$, we sampled their values from 0 to 270 degrees at
a step of $90^{\circ}$. The orbital propagation was carried out over a
time span of 120 years by means of the semi-analytical orbital
propagator FOP (Fast Orbit Propagator, see \cite{Ans, Rossi09} for
details), which accounts for the effects of $5\times 5$ geopotential,
SRP, lunisolar perturbations and atmospheric drag (below 1500 km of
altitude). We considered two values for the area-to-mass ratio:
$A/m=0.012$ m$^2/$kg, selected as a reference value for typical intact
objects in LEO, and $A/m=1$ m$^2/$kg, a representative value for a
small satellite equipped with an area augmentation device, as a solar
sail \cite{ColomboIAC}. The initial epoch for propagation was set to
21 June 2020. To catch the general behavior of the dynamics, we built
a set of maps showing the maximum eccentricity computed over the
propagation time span and the corresponding lifetime as a function of
the initial inclination and eccentricity, for each given semi-major
axis in the grid and for different $\Omega-\omega$ combinations. A
detailed analysis of the cartography of the LEO region has been
already presented by the authors in \cite{AlessiSD, AlessiIAC,
  SchettinoIAC, AlessiMNRAS, AlessiCMDA}. A larger set of dynamical
maps of LEO orbits can be found on the ReDSHIFT
website\footnote{http://redshift-h2020.eu/}.

In this work, the information gathered from such maps is used to
assess the possibility of a reentry strategy from a given LEO.  

\section{Single-burn strategy based on the predefined grid}
\label{sec:strat1}

To characterize the most suitable disposal maneuver to reenter we need
to consider that a limited maximum $\Delta v$ can be applied for an
impulsive maneuver, depending on the remaining on-board
propellant. The Gauss planetary equations (see, e.g., \cite{vallado})
can be applied to obtain a first guess of the achievable displacements
$(\Delta a, \Delta e, \Delta i)$, with a given
$\Delta \mathbf{v} = (\Delta v_r, \Delta v_t, \Delta v_h)$, where the
subscripts $r, t, h$ refer to the radial, transversal, out-of-plane
component, respectively. We have:
\begin{eqnarray}\nonumber
\Delta a &=& 2\frac{e\sin f}{n\sqrt{1-e^2}}\Delta v_r+2\frac{(1+e\cos f)}{n\sqrt{1-e^2}}\Delta v_t\\\label{eq:gauss_LEO}
\Delta e &=&\frac{\sqrt{1-e^2}\sin f}{na}\Delta v_r+\sqrt{1-e^2}\frac{\cos f +\cos E}{na}\Delta v_t\\\nonumber
\Delta i &=& \frac{r}{h}\cos u\Delta v_h\,,
\end{eqnarray}
where $f$ is the true anomaly, $E$ the eccentric anomaly, $u$ the
argument of latitude, $r$ the radius, $h$ the angular momentum, $n$
the mean motion.

The strategy implemented and described in the following consists in
applying an impulsive maneuver at the point of intersection of two
orbits: one which does not reenter in a given maximum time span and
another which does. Both departure and target orbits are taken from
the computed maps, meaning that the set of possible solutions is
constrained by the grid defined in Table \ref{grid}. Starting from a
given departure orbit which would not reenter in the selected time
span (e.g., 25 years), all the other initial conditions available in
the grid are explored, looking for those with a lifetime lower than
the desired threshold and such that the difference in $(a,e,i)$
corresponds to a total velocity change within the given maximum
$\Delta v-$budget.  

\begin{figure}
\begin{center}
\includegraphics[width=0.7\textwidth]{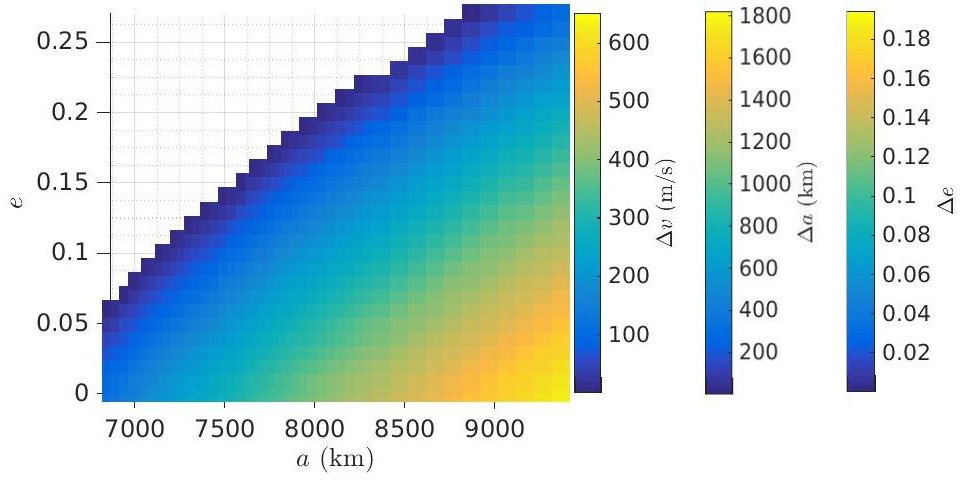}
\caption{$\Delta v$ (m$/$s), $\Delta a$ (km) and $\Delta e$ required
  to reenter by lowering the perigee altitude down to 80 km, as a
  function of initial ($a$, $e$). The change in $a$ and $e$ have been
  obtained from Eqs. (\ref{eq:gauss_LEO}) assuming a tangential
  maneuver.} \label{fig:dv_direct_reentry_LEO}
\end{center}
\end{figure}

\begin{figure}[h!]
\begin{center}
	\includegraphics[width=0.48\textwidth]{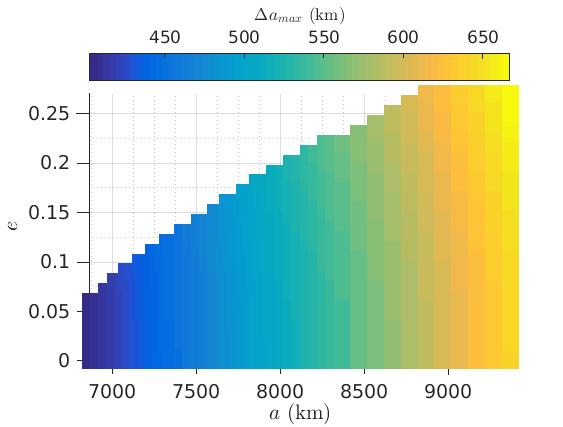} \includegraphics[width=0.48\textwidth]{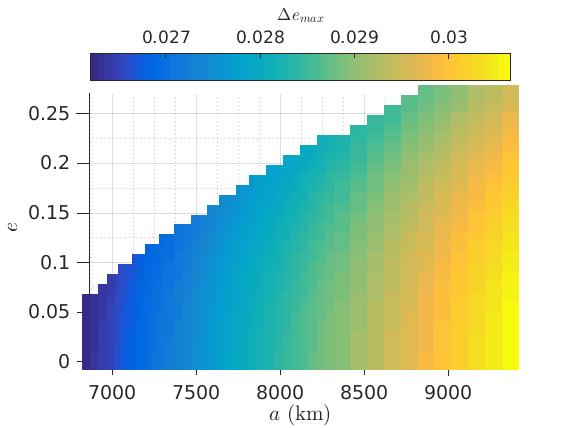}
\end{center}
\caption{Maximum change in $a$ (left; in km) and $e$ (right)
  achievable with a $\Delta v$ of 100 m$/$s, as a function of the
  initial $a$ and $e$.}
   \label{fig:deltaae_LEO}
\end{figure}

To assess the magnitude of the impulsive burn required to reenter
directly to the Earth, we look for the $\Delta v$ needed to lower the
altitude down to 80 km, following \cite{K06}. The required $\Delta v$
is shown as a function of $a$ and $e$ in Fig.
\ref{fig:dv_direct_reentry_LEO}, where we also show the color bar for
the corresponding $\Delta a$ and $\Delta e$, derived from
Eqs. (\ref{eq:gauss_LEO}) assuming a tangential maneuver.
In the case that the maximum $\Delta v$ available on-board is of 100
m$/$s, a plausible value for the maneuver budget, the corresponding
maximum variations achievable in $a$ and $e$ are shown in
Fig. \ref{fig:deltaae_LEO}.

\begin{figure}[h!]
\begin{center}
	\includegraphics[width=0.45\textwidth]{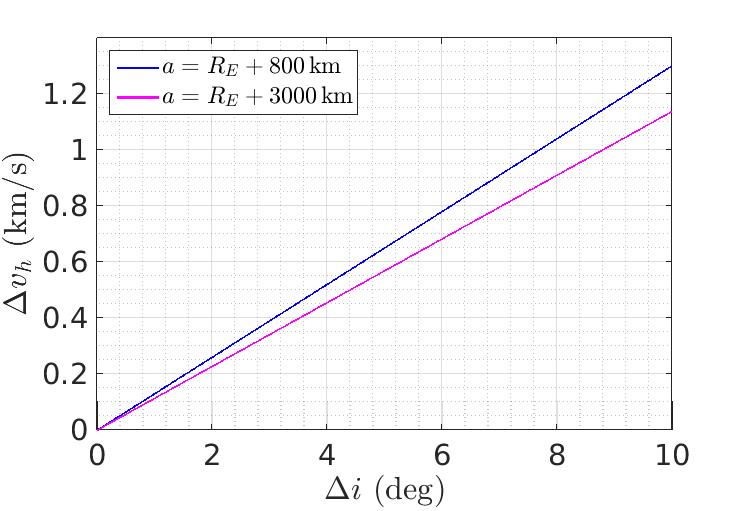} \includegraphics[width=0.45\textwidth]{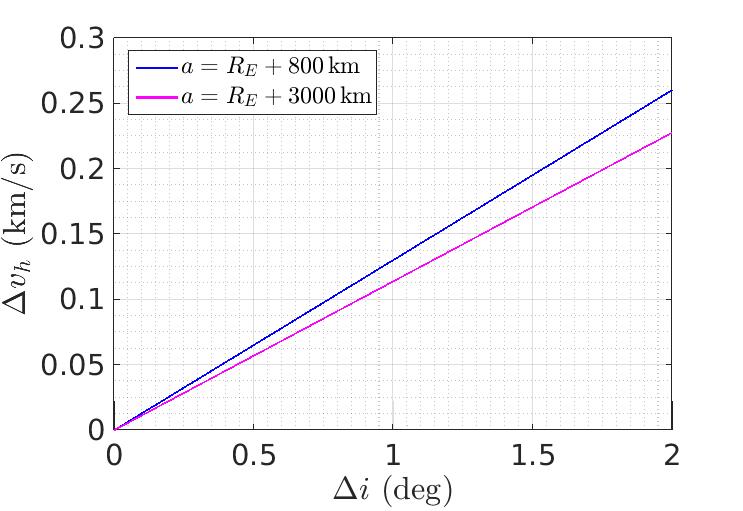}
\end{center}
\caption{Cost (km/s) required to change the inclination (deg) of a
  circular orbit at two different altitudes: 800 km in blue, 3000 km
  in magenta, assuming that the maneuver is applied at the node. On
  the right, a close-up up to $\Delta i =2^{\circ}$.}
   \label{fig:dvdi}
\end{figure}

Concerning the maximum allowed change in inclination, the third of
Eqs. (\ref{eq:gauss_LEO}) can be applied by assuming that the maneuver
provides only a variation in $i$, leaving the other orbital elements
unchanged. This is shown in Fig. \ref{fig:dvdi} -- left, where we display
the cost of a plane change maneuver for two circular orbits at
$a=R_E+800\,$km and $a=R_E+3000\,$km, respectively, in the case that
the maneuver is applied at the node. Fig. \ref{fig:dvdi} -- right
shows a close-up for $\Delta i$ up to $2^{\circ}$. The figure points out that
the required $\Delta v$ to provide a change in inclination of
$\Delta i=1^{\circ}$ is of the order of 100 m/s, almost independently
from the initial altitude. To achieve a higher $\Delta i$, a
considerable $\Delta v$ should be available: for example, the cost to
allow a change in inclination by $6^{\circ}$ should be as high as 700
m$/$s for an altitude of $3000\,$km and 800 m$/$s for an altitude of
$800\,$km. We also recall that the grid in inclination was set at a
step of $2^{\circ}$.

Notice that the Gauss equations help to filter the amount of data to
be examined.  After this filtering, before computing the actual
maneuver required to move from one orbit to the other, the implemented
algorithm checks if the two orbits intersect first in projection,
i.e., if the pericenter radius of the largest orbit is lower than the
apocenter radius of the smallest orbit, and, if it is the case,
computes the points of intersection by means of the procedure
explained in \cite{G05}. At these points, the $\Delta v$ is finally
computed by transforming the orbital elements into Cartesian
coordinates. In this way, we take into account the change in velocity
due to a possible change of all the orbital elements.

\subsection{First results: low area-to-mass ratio}
\label{sec:first_res}

\begin{figure}[h!]
\begin{center}
  \includegraphics[width=0.32\textwidth]{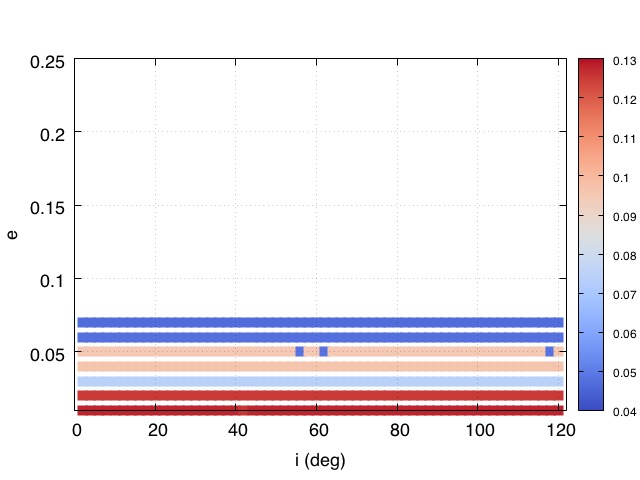} \includegraphics[width=0.32\textwidth]{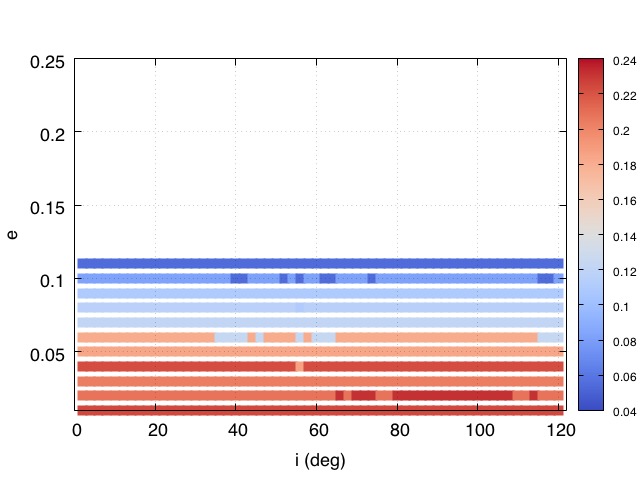} \includegraphics[width=0.32\textwidth]{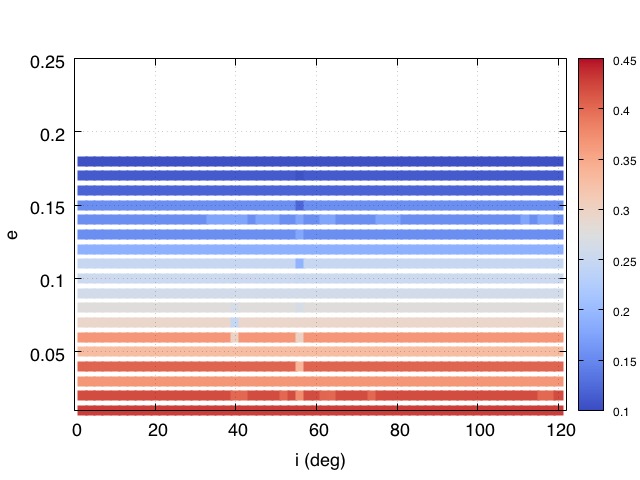}
  \includegraphics[width=0.32\textwidth]{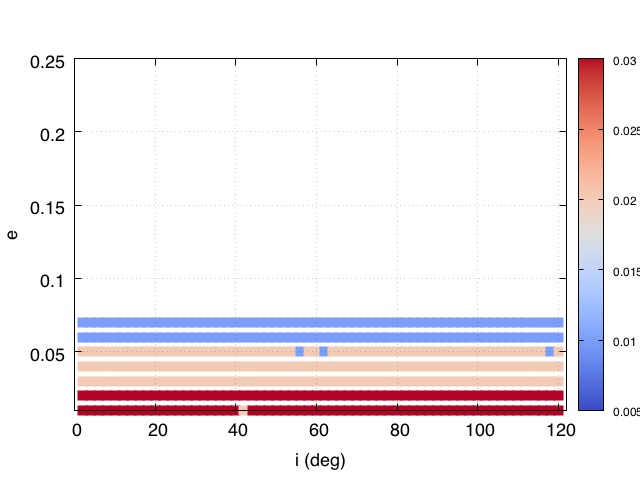} \includegraphics[width=0.32\textwidth]{interp_dvbest_all50y_26.jpg} \includegraphics[width=0.32\textwidth]{interp_dvbest_all50y_51.jpg}
  \includegraphics[width=0.32\textwidth]{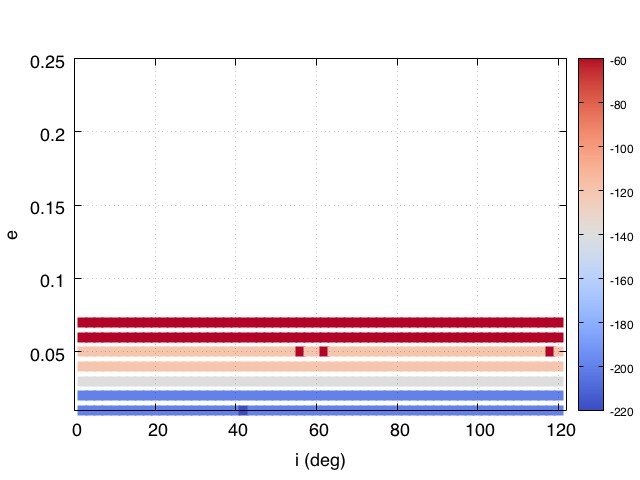} \includegraphics[width=0.32\textwidth]{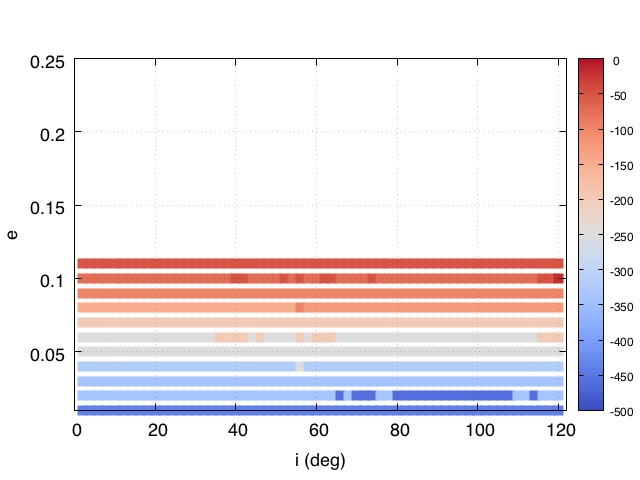} \includegraphics[width=0.32\textwidth]{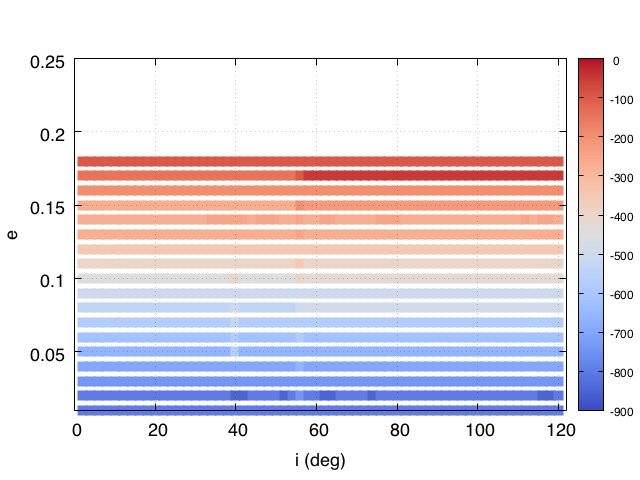}
\end{center}
    \caption{For $A/m=0.012\,$ m$^2/$kg, as a function of initial
      $(i,\,e)$ in the LEO region, examples of the ``minimum
      cost-reentry solutions'' computed with a single-burn maneuver
      for three values of the initial semi-major axis:
      $a=R_E+960\,$km (first column), $a=R_E+1300\,$km (second column)
      and $a=R_E+2100\,$km (third column). The color bar shows the
      $\Delta v$ computed (first row, units are m/s), the corresponding $\Delta e$
      (second row) and $\Delta a$ (third row, units are km).}
   \label{fig:first_res}
\end{figure}

As an example of the single-burn strategy described above, we compute
the minimum $\Delta v$ required for each initial orbit to reenter
within a 50 years residual lifetime, in the case of
$A/m=0.012\,$m$^2/$kg.  The generous 50 years limit is considered to
overcome the constraint imposed by the usage of the grid, and to
obtain some reentry options also for high values of semi-major
axis. The results are shown in Fig. \ref{fig:first_res} -- top row for
3 initial semi-major axes, $a=R_E+(960,\,1300,\,2100)\,$km, as a
function of the initial $(i,\,e)$.  Note the uniform behavior found,
except for isolated points (e.g., those located around
$i=40^{\circ}, 56^{\circ},64^{\circ},116^{\circ}$) corresponding to
the dynamical resonances, described in \cite{AlessiCMDA}. These
resonances are associated with perturbations different from the
atmospheric drag, and take place in narrow regions of the $(a,e,i)$
space. For a typical value of area-to-mass ratio they can reduce the
$\Delta v$ needed to reenter by an amount of the order of tens of m/s,
only if exploited along with the drag.

In general, the maneuver computed with the implemented strategy aims
at lowering the semi-major axis and increasing the eccentricity. When
a perturbation is exploited, these changes are smaller. This is shown
in Fig. \ref{fig:first_res} as well: the second row shows the
$\Delta e$ associated to the minimum required $\Delta v$ for the three
selected values of semi-major axis, while the third row shows the
corresponding $\Delta a$. An inclination change is, instead, almost
never chosen by the procedure, except for very specific cases. For
this methodology, the orbital elements that can be considered as a
target are the ones in the grid defined in Table \ref{grid}, that is,
we explore a discrete set of final conditions. In particular, the
inclination step is $\Delta i=2^{\circ}$, which limits the
applicability of the method, since a maneuver of $2^{\circ}$ in
inclination is, in general, above the threshold of the available
$\Delta v$. Considering a finer grid, i.e., $\Delta i=0.5^{\circ}$, in
some cases an inclination maneuver might become more convenient than
the usual semi-major axis and eccentricity one, and the exploitation
of the resonant corridors would be more evident. A specific
exploration of these possibilities will be discussed further in Sec.
\ref{sec:life}.

Some qualitative remarks can be done by comparing our results far from
resonances with the ones shown in \cite{JetAl03}, where end-of-life
deorbiting strategies for typical LEO satellites are considered. In
Table \ref{Tab_Jan}, we recall the results shown in \cite{JetAl03}
about the highest initial orbital altitude from which it is possible
to dispose a given spacecraft in compliance with the 25-year rule,
accounting for either a maximum $\Delta v$ of 100 m/s or 200 m/s.
Given that the orbits they considered are not exactly circular, and
that no details on the area-to-mass values are provided, their results
can be considered consistent with our findings. This can be inferred
by looking to Fig. \ref{fig:first_res} -- first row. In particular,
looking to the second panel of Fig. \ref{fig:first_res}, which refers
to an initial orbit at $a=R_E+960$ km, we can see that our single-burn
procedure gives a $\Delta v\simeq 130$ m/s to reenter from an initial
quasi-circular orbit, while a less expensive maneuver is sufficient to
deorbit from slightly more elliptical orbits. The same is true if we
look to the third panel of Fig. \ref{fig:first_res}, which refers to
$a=R_E+1300\,$km; in this case, a $\Delta v\simeq 240$ m/s is required
to reenter from quasi-circular orbits.
 \begin{table}[htb!]
 \centering
 \begin{tabular}{lcc}
 \hline
 Spacecraft & $\Delta v = 100$ m/s & $\Delta v = 200$ m/s \\
 \hline
 Pathfinder & 980 km & 1370 km \\
 Munin & 910 km & 1290 km \\
 Safir-2 & 870 km & 1240 km \\
 Abrixas & 910 km & 1260 km \\
 IRS\_1C & 910 km & 1300 km \\
 2420 kg-spacecraft & 870 km & 1250 km \\
 \hline
 \end{tabular}
 \caption{Maximum initial orbital altitude from which the given spacecraft can comply with the 25-year rule (values from  \cite{JetAl03}).}
 \label{Tab_Jan}
 \end{table}

\subsection{First results: high area-to-mass ratio}
 
The results obtained in \cite{AlessiSD, AlessiIAC, AlessiCMDA} show
that in order to be compliant with the 25-year rule the exploitation
of a drag sail might be successful for quasi-circular orbits up to an
altitude of about 1050 km, irrespective of the initial value of
inclination. On the other hand, in order to properly exploit the SRP
perturbation (i.e., a solar sail) the corresponding resonant
inclination bands shall be targeted.  Hence, for satellites where a
reasonably sized solar/drag sail can be mounted, we can conceive a
deorbiting strategy of two phases: in the first one a relatively small
maneuver is performed to reach a realm where, in the second phase,
either the atmospheric drag or the solar radiation pressure can be
exploited to reenter by means of a passive sail.  For the
$\Delta v-$threshold imposed in this case, namely, 100 m$/$s, a sail
of $A/m=1$ m$^2/$kg, can be effective up to $a=R_E+3000$ km, also for
quasi-circular orbits, but only if the satellite is moving in the
vicinity of one of the inclination corridors at $i\approx 40^{\circ}$
and $i\approx 80^{\circ}$, corresponding to the two main SRP
resonances (see, e.g., \cite{AlessiMNRAS} for details).

\begin{figure}[h!]
\begin{center}
  \includegraphics[width=0.32\textwidth]{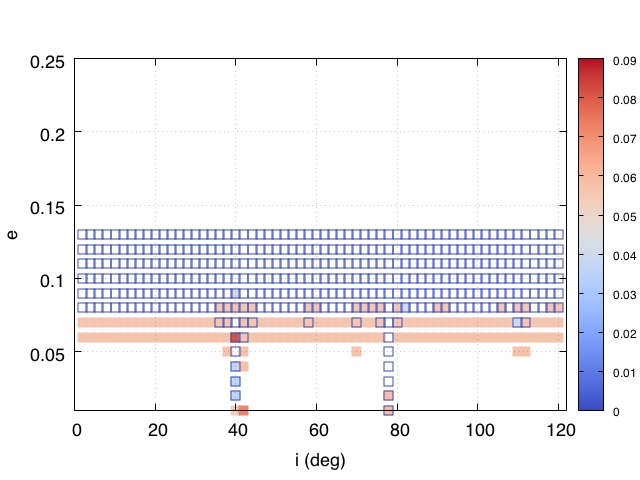} \includegraphics[width=0.32\textwidth]{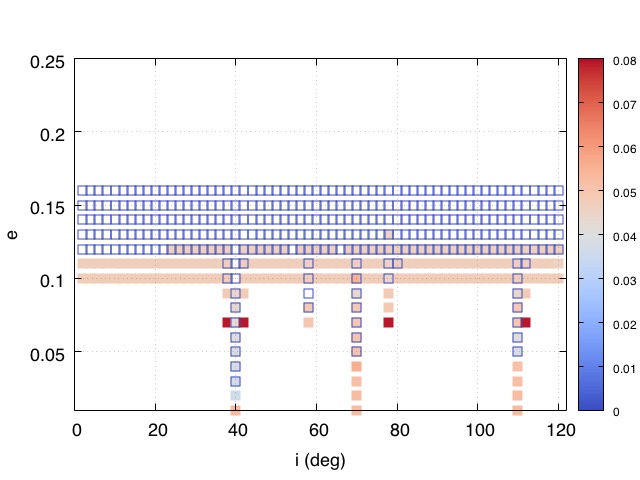} \includegraphics[width=0.32\textwidth]{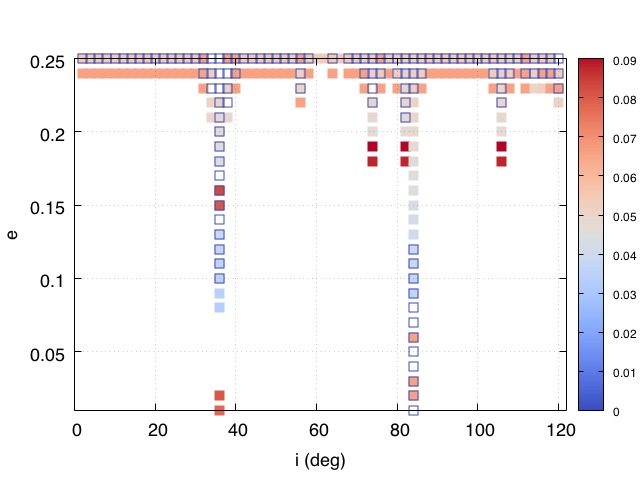}
  \includegraphics[width=0.32\textwidth]{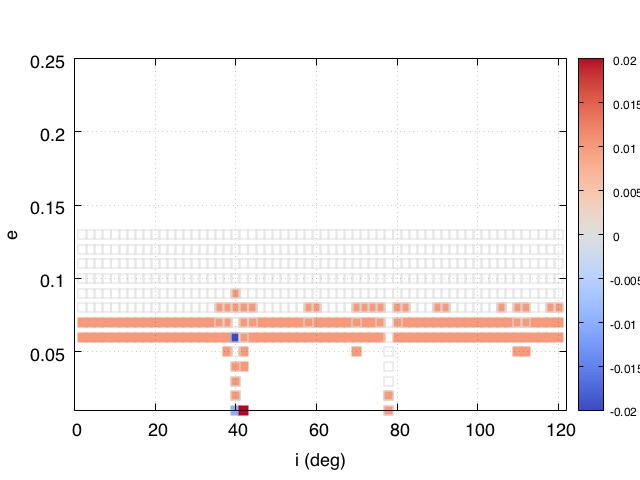} \includegraphics[width=0.32\textwidth]{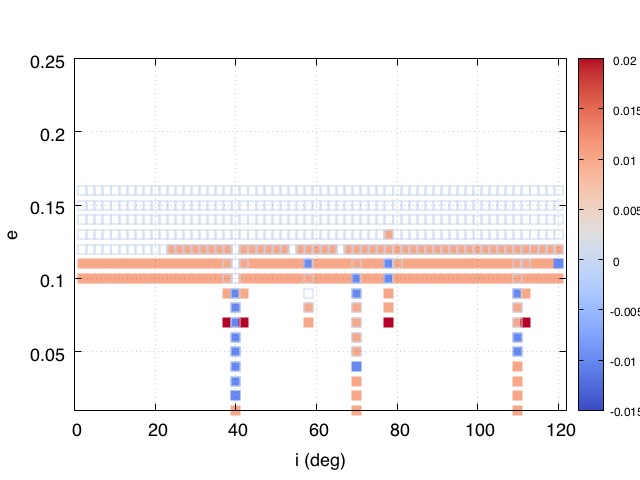} \includegraphics[width=0.32\textwidth]{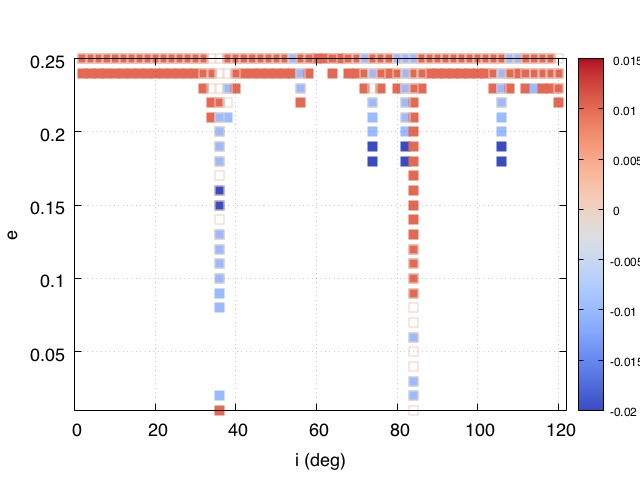}
  \includegraphics[width=0.32\textwidth]{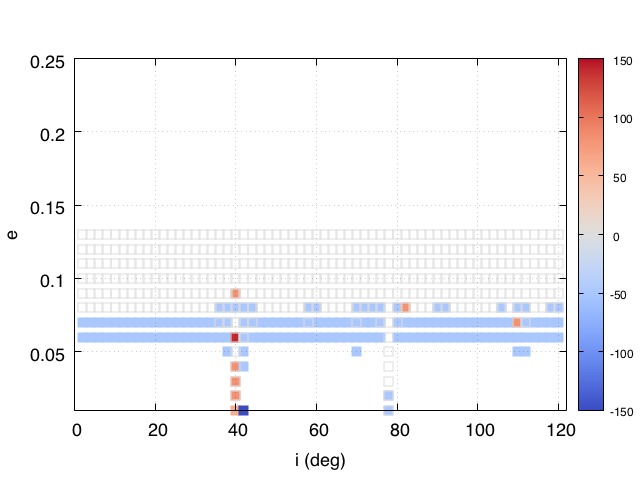} \includegraphics[width=0.32\textwidth]{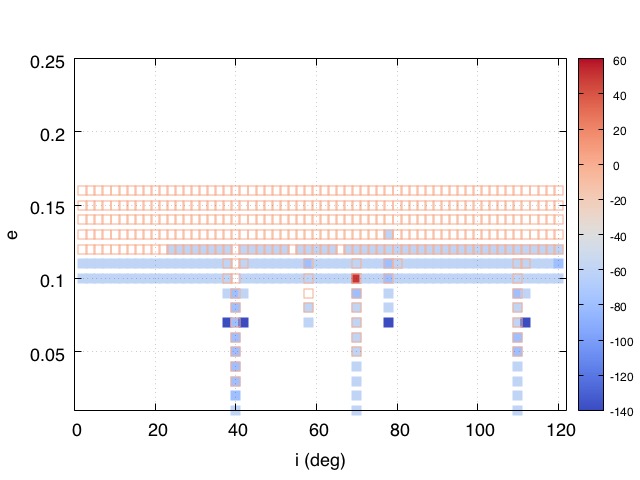} \includegraphics[width=0.32\textwidth]{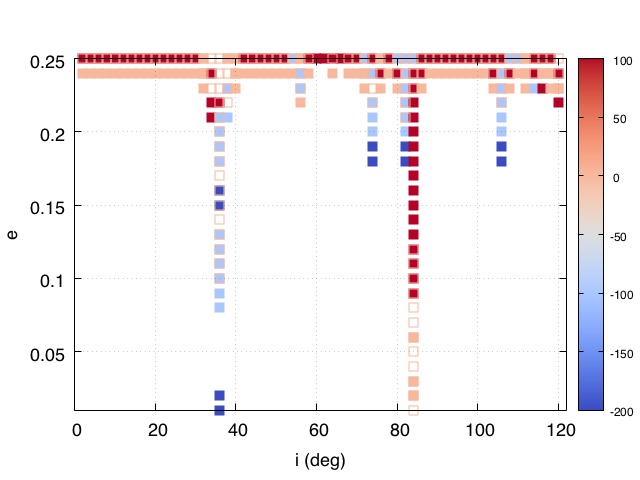}
\end{center}
    \caption{For $A/m=1\,$ m$^2/$kg, as a function of initial
      $(i,\,e)$ in the LEO region, examples of the ``minimum
      cost-reentry solutions'' for three values of the initial
      semi-major axis: $a=R_E+1300\,$km (first column),
      $a=R_E+1600\,$km (second column) and $a=R_E+2900\,$km (third
      column). The color bar shows the $\Delta v$ computed (first row,
      units are m/s), the corresponding $\Delta e$ (second row) and
      $\Delta a$ (third row, units are km).}
   \label{fig:first_res_high}
\end{figure}

As in Fig. \ref{fig:first_res}, in Fig. \ref{fig:first_res_high} --
first row we show the minimum $\Delta v$ computed to reenter in 25
years exploiting a drag or a solar sail (given the maximum threshold
of 100 m$/$s) in the $(i,\,e)$ space for three values of the initial
semi-major axis, $a=R_E+(1300,\,1600,\,2900)\,$km.
In the plots, white regions denote initial conditions for which there
does not exist a solution, i.e., the solutions computed are
characterized by a $\Delta v$ higher than the limit set. Empty squares
represent natural solutions which do not need any impulsive strategy
but only the usage of a sail, while colored squares refer to initial
conditions requiring both an impulsive maneuver and the exploitation
of a sail. Empty and colored squares may overlap, because we
considered all the possible $(\Omega, \omega)$ configurations of the
grid. In Fig.  \ref{fig:first_res_high} -- second row we show the
change in eccentricity corresponding to the minimum $\Delta v$ for the
same three values of semi-major axis, while in the third row we
present the corresponding variation in semi-major axis. For each case
considered, a change in inclination was never chosen by the procedure,
since it never corresponds to a minimum cost solution.
 
\section{Single-burn strategy based on the dynamical effects}
\label{sec:life}

The algorithm just described allows to compute, for a given initial
condition, various solutions, which can differ not only in the
$\Delta v$ required to reenter, and thus in the target orbital
elements, but also in the lifetime. In particular, as mentioned above,
being based on the grid discretization adopted in the mapping of the
LEO phase space, our results are step-wise continuous. Hence, for some
specific orbits, we cannot target a specific value of the residual
lifetime (hence of the required $\Delta v$), while the only
information available is whether the solution is compliant with the
25-year (or ``desired''-year) rule or not. Moreover, due to the steps
in $(a,e,i)$ adopted in the grid, the target conditions fulfilling the
25-year rule are also step-wise continuous, and thus, we can miss
important information. For instance, in the non-resonant case, it can
happen that for a given value of $a$ there is a value of $e$ in the
grid which corresponds to a 25-year reentry, but for the next point of
$a$ in the grid the eccentricity available in the grid is associated
with a lifetime slightly higher than 25 years, and thus the above
methodology discards such semi-major axis as a possible
target. Nevertheless, note that, once the algoritm is implemented and
tested, there is the possibility to conveniently refine the grid as a
future task.

In general, we can be interested in computing the optimal maneuver,
exploiting a maximum available $\Delta v$ and requiring also for a
maximum residual lifetime. In this case, the assumption that the
target orbit belongs to the grid may become too severe. For these
reasons, we have attempted to employ the numerical cartography
computed in LEO in a different way. The theoretical analysis performed
(see, e.g., \cite{AlessiSD, AlessiIAC, AlessiCMDA, AlessiMNRAS}) show
that, for a given semi-major axis and eccentricity of the initial
orbit, the dynamical instabilities, triggered by perturbations
different from the atmospheric drag, can play a role only at specific
values of inclination. In particular, the resonant dynamics is
effective within narrow corridors in inclination, limited by at most
2$^{\circ}$ around the resonant value of inclination. Outside these
regions, the computed lifetime and the maximum eccentricity achieved
during propagation do not depend significantly on the initial
inclination. This information can be exploited in two ways, in order
to check if a reentry can be achieved in a given time:
\begin{itemize}
\item if the inclination of the departure orbit does not belong to the
  resonant corridor and the $\Delta v-$budget does not allow to reach
  it, then the initial reentry state is defined only in terms of
  semi-major axis and eccentricity, and the transfer will be driven by
  the atmospheric drag;
\item if a resonant value of inclination can be targeted with the
  available propellant on-board, then the impulsive strategy aims at
  changing semi-major axis, eccentricity and inclination, and the
  transfer will be driven by the atmospheric drag in combination with
  another perturbation.
\end{itemize}

For a given departure orbit, we can compute a set of displacements in
terms of semi-major axis, eccentricity and inclination, say
$(\delta a,\delta e,\delta i)$, defined as the difference, in terms of
$(a,e,i)$, between the initial conditions of the departure orbit and
all the possible target conditions associated to a reentry in the
desired time. Such conditions are determined by either the effect of
the atmospheric drag or the atmospheric drag together with another
perturbation.  On the other hand, the values
$(\Delta a,\Delta e,\Delta i)$ defined in Eqs. (\ref{eq:gauss_LEO})
provide an upper limit to the possible displacement achievable with a
given $\Delta v$. Thus, comparing the set
$(\Delta a,\Delta e,\Delta i)$ with each set
$(\delta a,\delta e,\delta i)$, we can find out if there exists at
least one reentry solution given the departure orbit, the
$\Delta v-$budget and the area-to-mass ratio.

For a given departure orbit, we can compute:
\begin{itemize}
\item the target values $(\delta a,\delta e)_{d}$ associated with a
  reentry assisted only by the drag in the desired time span;
\item the target values $(\delta a,\delta e, \delta i)_{p}$ associated
  with a reentry assisted by the drag plus another dynamical
  perturbation in the desired time span.
\end{itemize}
So, we may encounter the following situations:
\begin{itemize}
\item $\delta a_d \leq \Delta a$ and $\delta e_{d} \leq \Delta e$ for
  one or more target conditions, that is, the reentry is feasible by
  changing semi-major axis and eccentricity;
\item $\delta a_p \leq \Delta a$ and $\delta e_{p} \leq \Delta e$ and
  $\delta i_{p} \leq \Delta i$, for one or more target conditions,
  that is, the reentry is feasible by changing semi-major axis,
  eccentricity and inclination;
\item $\delta a_d > \Delta a$ or $\delta e_{d} > \Delta e$, that is,
  we cannot reenter only exploiting the effect of the atmospheric
  drag;
\item $\delta a_p > \Delta a$ or $\delta e_{p} >\Delta e$ or
  $\delta i_{p} > \Delta i$, that is, the reentry cannot be assisted
  by perturbations different from drag.
\end{itemize}
Note that, if more reentry conditions can be targeted, the selection
can be made following different criteria. In particular, one can
select the less expensive strategy in terms of $\Delta v$ or exhaust
all the available propellant to speed-up the reentry and passivate the
spacecraft.

\subsection{Target conditions: non-resonant case}\label{sec:i_nonres}

As mentioned above, the definition of the target conditions depends on
whether the reentry is driven by the atmospheric drag alone or in
combination with another perturbation. That is, it depends if we cannot
exploit a resonant condition or if it becomes possible.

\begin{figure}
\begin{center}
	\includegraphics[width=0.49\textwidth]{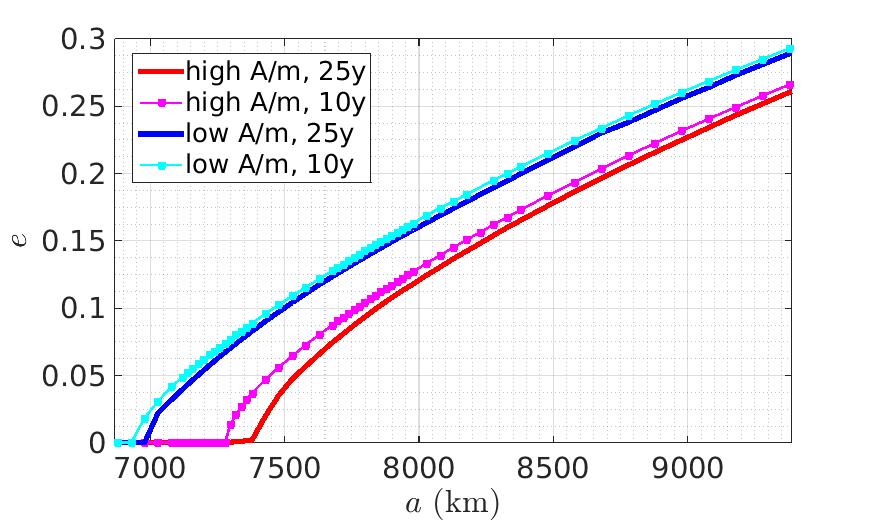} \includegraphics[width=0.49\textwidth]{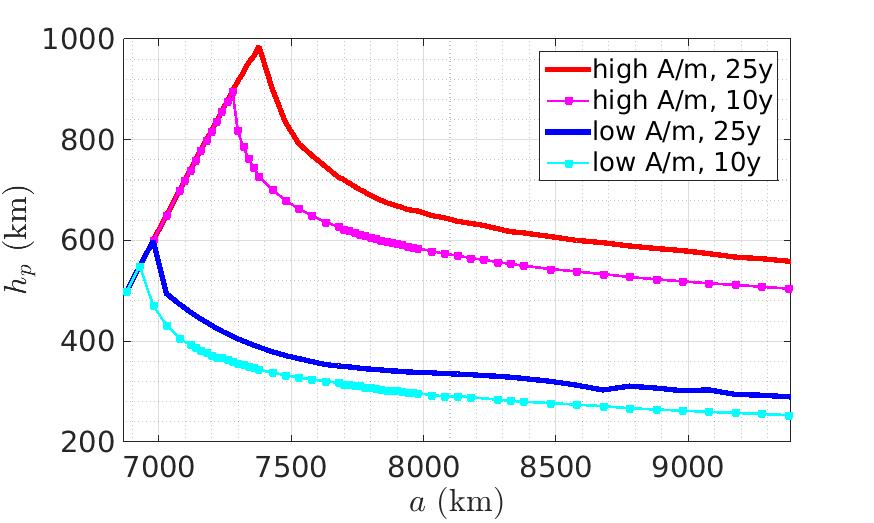}
\end{center}
\caption{On the left: initial $e$ required to reenter in 25 years
  (solid line) or 10 years (squares) for each $a$ of the grid at
  initial $i=10^{\circ}$, for both $A/m=0.012$ m$^2/$kg and $A/m=1$
  m$^2/$kg. On the right: initial pericenter altitude, $h_p$, required
  to reenter in 25 years (solid line) or 10 years (squares) for each
  $a$ of the grid at initial $i=10^{\circ}$, for both $A/m=0.012$
  m$^2/$kg and $A/m=1$ m$^2/$kg.}
   \label{figAA_1}
\end{figure}

In the former case, investigating the maps and considering also the
frequency characterization of the eccentricity evolution described in
\cite{SchettinoIAC}, we selected $i=10^{\circ}$ as a reference value
of initial inclination corresponding to a ``non-resonant'' dynamical
behavior. We selected two possible values of the residual lifetime: 25
years and the more compelling value of 10 years. Then, for each
initial semi-major axis of the grid we identified the required
eccentricity to reenter in 25 or 10 years. The results for both
area-to-mass ratios are shown in Fig.  \ref{figAA_1}, where we show,
on the left, the required $e$ and, on the right, the pericenter
altitude $h_p$ to reenter as a function of the initial $a$. These
$a-e$ configurations are the target conditions for the disposal
strategy.

Given our assumption on the solar flux\footnote{We assumed an
  exospheric temperature of 1000 K and a variable solar flux at 2800
  MHz.}, in the case of low $A/m$ ratio, the atmospheric drag turns out
to be effective in driving a reentry from quasi-circular orbits only
up to altitudes of about 600 km for a residual lifetime of 25 years
and of 550 km for a lifetime of 10 years, while at higher altitudes a
reentry within 25 years or less becomes feasible only if the initial
orbit is gradually more eccentric.  For $A/m=1$ m$^2/$kg, the drag
dominates the dynamics up to an altitude of about 1000 km and a
reentry within 25 years can be achieved. A reentry within 10 years,
instead, is feasible if the pericenter altitude is below 900 km.

\subsection{Target conditions: resonant case}\label{sec:i_res}

\begin{figure}
\begin{center}
	\includegraphics[width=0.6\textwidth]{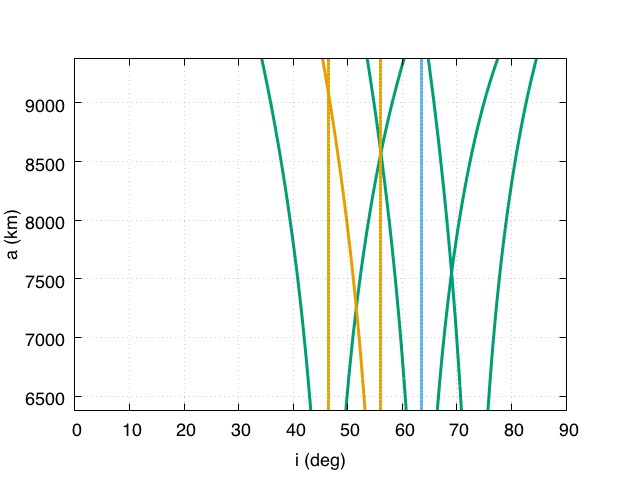}
\end{center}
\caption{As a function of the initial inclination (deg) and semi-major
  axis (km), we show the location of the resonances found to play a
  role in the LEO region. The color of the curves describe the nature
  of the resonance: green for SRP, cyan for 5th degree zonal harmonic,
  ocher for lunisolar perturbations. The curves have been computed for
  $e=0.01$.}
   \label{fig:res}
\end{figure}

In the resonant case, instead, the target conditions are $a-e-i$
configurations, which correspond to a resonance involving the rate of
precession of the ascending node $\dot\Omega$ and of the argument of
pericenter $\dot\omega$. Solar radiation pressure, lunisolar
perturbations and the 5th zonal harmonics produce a long-term
variation in eccentricity, which becomes quasi-secular when a
well-defined combination of $\dot\Omega$ and $\dot\omega$ tends to
zero. In LEO, assuming the two values of $A/m$ adopted in this work,
the rate of $\Omega$ and $\omega$ can be approximated considering only
the effect of the oblateness of the Earth. As a consequence, the
growth of eccentricity and the corresponding reduction in lifetime can
be expressed as a function of $(a,e,i)$. Thus, given the semi-major
axis and eccentricity, the resonances are arranged along inclination
curves. An example of the resonant curves as a function of $i$ and $a$
fixing $e=0.01$ is shown in Fig. \ref{fig:res}: the green curves refer
to resonances due to SRP, the cyan curve to 5th degree zonal harmonic
and the ocher curves to lunisolar perturbations. More details can be
found in \cite{AlessiSD, AlessiCMDA, AlessiMNRAS}. In order to benefit
from the dynamical perturbation responsible of the resonance and to
facilitate a reentry, the initial orbit must lay within one of the
highlighted resonant corridors.  Associated with such values of
inclination, the required eccentricity to reenter within 25 years can
be significantly lower than at the nearby inclinations.

\begin{figure}
\begin{center}
	\includegraphics[width=0.7\textwidth]{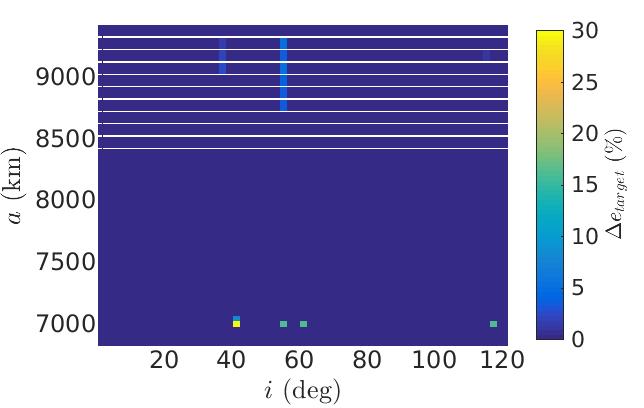}
\end{center}
\caption{Relative difference $\Delta e_{target}$ (in \%) between the
  eccentricity needed to reenter in 25 years at $i=10^{\circ}$
  ($e_{d}$) and the eccentricity of the target orbit in case of a
  resonance which assists the reentry ($e_{p}$), for $A/m=0.012$
  m$^2/$kg, as a function of the initial $i$ and $a$.}
   \label{figAA_2}
\end{figure}

The initial $a$ and $i$ where the resonant behavior allows for a lower
target eccentricity are highlighted in Fig.  \ref{figAA_2} for the
case $A/m=0.012$ m$^2/$kg. The color bar refers to the relative
difference $\Delta e_{target}=(e_{d}-e_{p})/e_{d}$ between the
reference target eccentricity at $i=10^{\circ}$, $e_{d}$, and the
target eccentricity required at a resonant inclination, $e_{p}$.
As expected for the low value of area-to-mass ratio, most of the
figure is dark blue, except a few specific points where resonances are
effective in reducing the target eccentricity. Writing the disturbing
function due to a given perturbation as a sinusoidal term of argument
$\psi$ (see \cite{AlessiMNRAS} for more details), the observed
resonances correspond to (refer also to Fig. \ref{fig:res}):
\begin{itemize}
\item $\dot\psi=\dot\Omega+\dot\omega-\dot\lambda_S\simeq 0$, around
  $i=40^{\circ}$ and $i=116^{\circ}$, associated to SRP (see, e.g.,
  \cite{Hughes77}), where $\lambda_S$ is the longitude of the Sun with
  respect to the ecliptic plane;
\item $\dot\psi=\dot\Omega+2\dot\omega$ at $i=56^{\circ}$, which is
  the well-known lunisolar gravitational resonance; moreover, for
  altitudes above $h=2000$ km, at the same inclinations are effective
  also two resonances due to SRP corresponding to
  $\dot\psi=\dot\omega-\dot\lambda_S\simeq 0$ and
  $\dot\psi=\dot\Omega+\dot\omega+\dot\lambda_S\simeq 0$, respectively
  (see, e.g., \cite{Hughes77});
\item $\dot\psi=\dot\omega\simeq 0$ at $i=63.4^{\circ}$, due to the
  doubly-averaged lunisolar gravitational perturbations and to the
  5th-degree zonal harmonics (see, e.g., \cite{Cook, Hughes80,
    Merson61, AlessiCMDA}).
\end{itemize}
Note that Fig. \ref{figAA_2} was obtained displaying the data
computed by the numerical cartography, and analogous results are
available for all the explored values of eccentricity. The figure
allows to conclude that in practice, when a reentry within 25 years is
required, the benefit due to resonances is limited in the case of low
$A/m$ ratio. This fact does not disagree with the results on the
global LEO dynamical mapping found by the authors and described, e.g.,
in \cite{AlessiSD, AlessiIAC, AlessiCMDA}. Indeed, we observed that
resonances could assist a reentry at specific inclinations, given
semi-major axis and eccentricity, by lowering the residual lifetime of
some tens of years, in combination with the atmospheric drag. The
observed decrease in lifetime was, however, compliant with the 25-year
rule only in a few cases for $A/m=0.012$ m$^2/$kg. 

\begin{figure}
\begin{center}
	\includegraphics[width=0.7\textwidth]{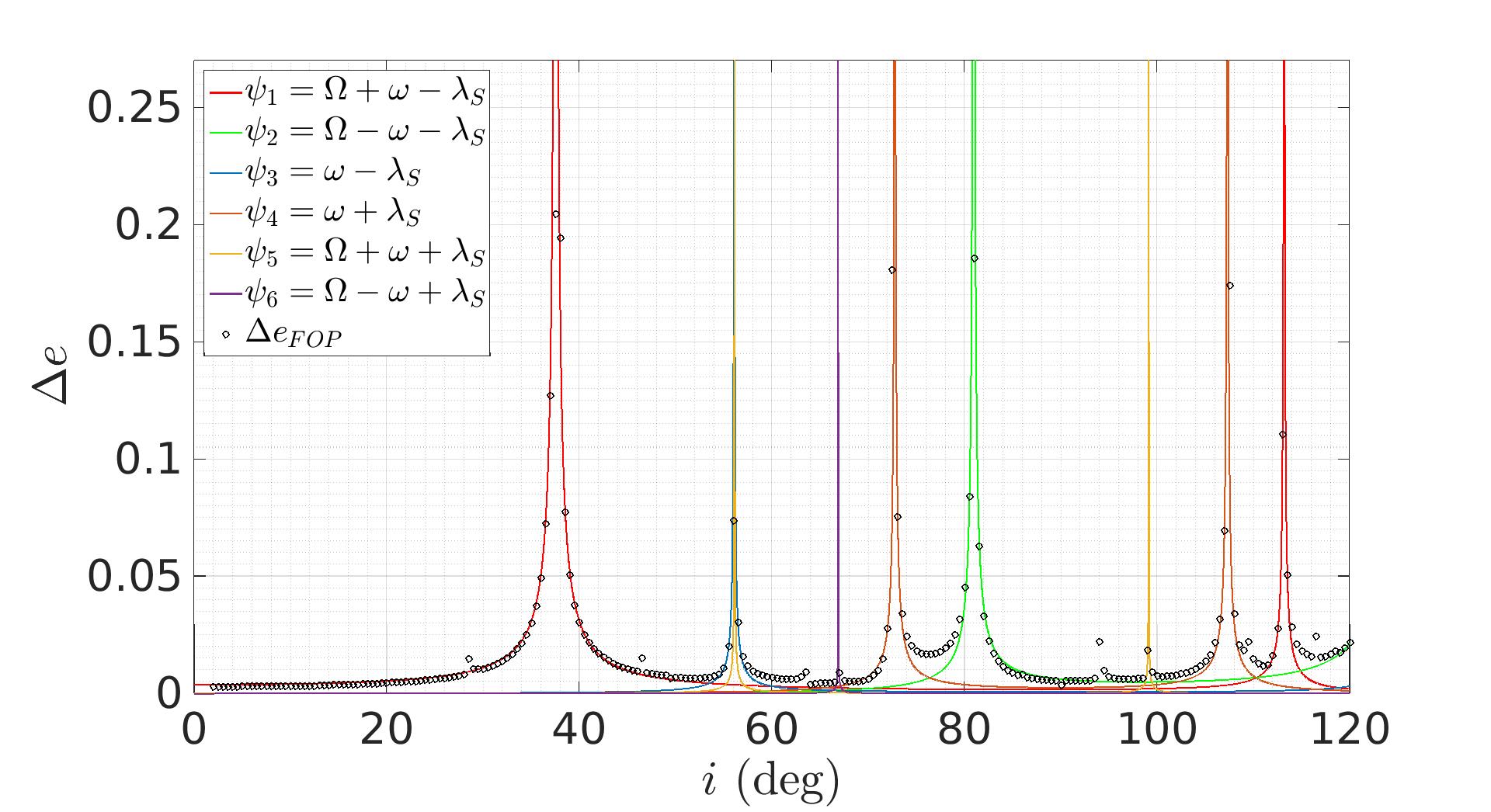}
\end{center}
\caption{Comparison between the maximum eccentricity variation,
  computed with Eq. (\ref{eq:SRP}), and the eccentricity variation
  provided by FOP, as a function of inclination, for the initial
  orbit: $a=R_E+2200\,$km, $e=0.001$, $\Omega=\omega=0^{\circ}$.}
   \label{figAA_SRP}
\end{figure}

In the case of $A/m=1$ m$^2/$kg, the approach is different. The target
$a-e-i$ conditions are not obtained from the numerical cartography,
but they can be computed analytically. Due to the high area-to-mass
ratio, the only resonances which matter are those associated with the
SRP. In this case, considering only the first order terms, the
disturbing function can be written as a sum of sinusoidal terms with
argument:
\begin{equation}
\psi = \alpha\Omega \pm\omega\pm \lambda_s \,\,\,\,\,\,(\alpha=0,1)\,,\nonumber
\end{equation}
corresponding to six different resonances, indexed as in Table
\ref{item_res} (see, e.g., \cite{Krivov}). In \cite{AlessiMNRAS} we
developed a simplified analytical theory which allows to compute the
supreme norm of the variation in eccentricity induced by the six first
order SRP resonances. The maximum eccentricity variation that can be
achieved at the resonance $j$ ($j=1,..6$) can be estimated as:
\begin{equation}
\Delta e_{SRP,j}=\left | \frac{3}{2}\,P\,C_R\,\frac{A}{m}\,\frac{\sqrt{1-e^2}}{na}\,\frac{T_j}{\psi_j}  \right |\,,
\label{eq:SRP}
\end{equation}
where $P$ is the solar radiation pressure, $C_R$ the reflectivity
coefficient, $n$ the mean motion of the spacecraft and the explicit
expression for $T_j$ is shown in Table \ref{item_res} (see, e.g.,
\cite{AlessiMNRAS}).
\begin{table}[h!]
\begin{center}
\begin{tabular}{ccc}
\hline
$j$ & $\psi_j$ & $T_j$ \\
\hline
1 & $\Omega +\omega -\lambda_S$ & $\cos^2 (\epsilon /2)\cos^2 (i/2)$ \\
2 & $\Omega -\omega -\lambda_S$ &  $\cos^2 (\epsilon /2)\sin^2 (i/2)$ \\
3 & $\omega -\lambda_S$ & $1/2 \sin(\epsilon) \sin (i)$ \\
4 & $\omega +\lambda_S$ & $-1/2 \sin(\epsilon) \sin (i)$ \\
5 & $\Omega +\omega +\lambda_S$ & $\sin^2 (\epsilon /2)\cos^2 (i/2)$ \\
6 & $\Omega -\omega +\lambda_S$ & $\sin^2 (\epsilon /2)\sin^2 (i/2)$ \\
\hline
\end{tabular}
\caption{For each first order resonance due to SRP ($j=1,..6$), we
  show the argument $\psi_j$ and the corresponding function $T_j$
  ($\epsilon$ refers to the obliquity of the
  ecliptic).} \label{item_res}
\end{center}
\end{table}
The comparison between the theoretical variation in
eccentricity, $\Delta e_{SRP,j}$, due to each resonance and the
eccentricity increment, $\Delta e_{FOP}$, computed over 120 years of
propagation with FOP, is shown in Fig. \ref{figAA_SRP} as a function
of the inclination, for the initial orbit: $a=R_E+2200\,$km,
$e=0.001$, $\Omega=\omega=0^{\circ}$. In this case, the step in the
inclination grid was set to $\Delta i=0.5^{\circ}$. This plot,
together with analogous ones for different values of $(a,e)$, confirms
a very good match between the theoretical and computed variation of
eccentricity, supporting the assumption that a first-order theory
describes accurately the dynamics induced by SRP.

\begin{figure}
\begin{center}
	\includegraphics[width=0.7\textwidth]{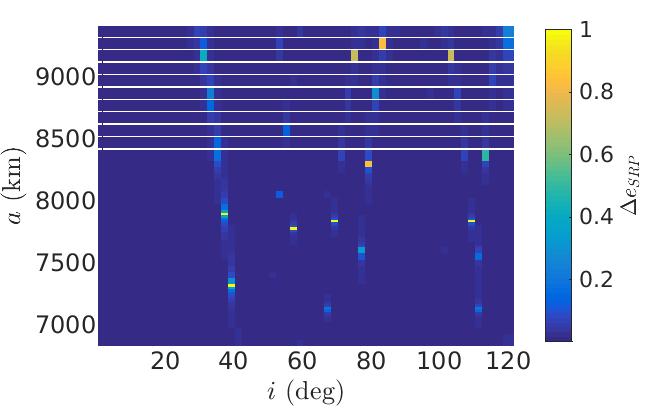}
\end{center}
\caption{Increase in eccentricity due to SRP as a function of the
  initial $a$ and $i$, for $e=0.01$ and $A/m=1$ m$^2/$kg.}
   \label{figAA_3}
\end{figure}

The above expression can be used to compute the total variation in
eccentricity due to SRP:
\begin{equation}\nonumber
\Delta e_{SRP}=\sum_{j=1}^6\Delta e_{SRP,j}\,. \nonumber
\end{equation}
First, the given $\Delta v$ provides a set $(a,e,i)$ of attainable
orbits that can be targeted starting from the initial orbit. Note that
the attainable set includes the initial condition itself, that is, it
includes all the possible $(a,e,i)$ achievable applying a maneuver at
most equal to the given $\Delta v$. Then, the reentry can be supported
by SRP if the $\Delta e_{SRP}$ associated to one of the target orbits of
the set ensures to reach the proper curve in Fig. \ref{figAA_1}.  As
an example, in Fig. \ref{figAA_3} we show, as a function of the
initial $i$ and $a$, the displacement $\Delta e_{SRP}$ induced by SRP,
fixing the initial $e=0.01$. The initial orbits where SRP alone is
capable of triggering the reentry can be identified comparing the
computed value of $\Delta e_{SRP}$ with the eccentricity required to
reenter in 25 years, corresponding to the red line in
Fig. \ref{figAA_1} -- left. For example, for $A/m=1$ m$^2/$kg
Fig. \ref{figAA_1} -- left shows that at an altitude of 2200 km the
eccentricity needs to be, at least, $e=0.18$ to reenter within 25 years;
thus, as shown in Fig. \ref{figAA_3}, around the resonant inclinations
the growth of eccentricity induced by SRP alone ensures reentry.

\subsection{Results: low area-to-mass ratio}
\label{sec:res}

To show the possible outputs of the described disposal strategy in the
case of $A/m=0.012\,$m$^2/$kg, we consider two test cases whose
initial orbital elements are detailed in Table \ref{Tab:case_low}.
\begin{table}[htb!]
\centering
\begin{tabular}{ccccccc}
\hline
 & $a$ (km) & $e$ & $h_p$ (km) & $i$ ($^{\circ}$) & $\Omega$ ($^{\circ}$) & $\omega$ ($^\circ$) \\
\hline
Case 1 & 7100 & 0.02 & 580 & 40.8 & 90 & 0 \\
Case 2 & 7100 & 0.02 & 580 & 41.8 & 90 & 0 \\
\hline
\end{tabular}
\caption{Initial orbits selected as a reference case for
  $A/m=0.012\,$m$^2/$kg: semi-major axis, eccentricity, altitude of
  the perigee, inclination, right ascension of the ascending node,
  argument of perigee.}
\label{Tab:case_low}
\end{table}
The two initial orbits differ only by 1$^{\circ}$ in inclination. We
recall that at this altitude the resonant inclination associated to
the dominant term $\psi_1$ due to SRP is $i_{res}=41.8^{\circ}$. Thus,
the second case refers to an orbit with initial inclination
corresponding exactly to a resonant value, while the first case
considers an orbit lying only $1^{\circ}$ in inclination next to a
resonant corridor. We assume to have an available maximum $\Delta v$
of 80 m/s. For this altitude and eccentricity, the tangential maneuver
to be applied at the apogee in order to lower the perigee down to 120
km would be of 130 m/s, thus a direct reentry is not allowed. In the
first case, the minimum cost solution obtained consists in applying
$\Delta v_t=-33.6$ m/s at the apogee which corresponds to
$\Delta a=-62$ km and $\Delta e=0.009$.  The perigee altitude of the
target orbit is $h_p=455.7$ km: at this altitude the effect of
atmospheric drag guarantees a reentry in 25 years.

In the second case, instead, a resonant solution due to SRP can be
exploited: with a total $\Delta v=28.2$ m/s, lower than the previous
case, it is possible to change the semi-major axis by $\Delta a=-22$
km, the eccentricity by $\Delta e=0.003$ and the inclination by
$\Delta i=0.0035^{\circ}$, targeting a new orbit where SRP can be
exploited in order to achieve a reentry in 25 years. This result
confirms that we can take advantage from a resonance to facilitate a
reentry, requiring a lower $\Delta v-$budget. Note, however, that the
inclination range where the perturbation is effective is very narrow
and that the gain in terms of the required $\Delta v$ is also limited.

A further interesting information concerns the required propellant mass to
perform the computed maneuver.  Following \cite{K06}, the propellant
mass required to perform the maneuver is given by:
\begin{equation}
\frac{m_0-m_f}{m_0}=1-\exp \left (-\frac{\Delta v}{w_e} \right )\,,
\end{equation}
where $w_e$ is the exhaust velocity. In Table \ref{Tab_Klink}, we
compare the results we have obtained with the ones shown in \cite{K06}
- Chapter 6, for some representative altitudes. In columns 2--3 we
show the maneuver and the propellant mass fraction, $\Delta v_K$ and
$(\Delta m/m_0)_K$, respectively, reported in \cite{K06} for a direct
deorbiting from a circular orbit at altitude $H$ to an elliptical
orbit with a perigee altitude $H_p=80$ km, assuming $w_e=2747$ m/s. In
columns 4--5 we show the maneuver and the propellant mass fraction,
respectively, arising from our analysis for an initial quasi-circular
orbit ($e=10^{-3}$), considering the minimum cost deorbiting option
complying with the 25-year rule and the $A/m=0.012\,$m$^2/$kg
ratio. The initial inclination has been set at $i=10^{\circ}$, far
from any resonance. The same $w_e$ is applied to evaluate our
outcome. From the table, it is apparent the well-known benefit of
choosing a 25-year disposal reentry rather than a direct one, in terms
of $\Delta v$ cost and mass consumption.
  \begin{table}[htb!]
\centering
\begin{tabular}{ccccc}
\hline
$H$ (km) & $\Delta v_{K}$ (m/s) & $(\Delta m/m_0)_K$ (\%) & $\Delta v$ (m/s) & $\Delta m/m_0$ (\%) \\
\hline
800 & 199.4 & 7.0 & 75.0 &  2.7 \\
900 & 224.3 & 7.8 & 123.0 & 4.4 \\
1000 & 248.6 & 8.6 & 135.0 & 4.8 \\
1100 & 272.3 & 9.4 & 162.0 & 5.7  \\
1200 & 295.4 & 10.2 & 187.0 & 6.6 \\
1300 & 317.9 & 10.9 & 211.5 &  7.4 \\
1400 & 339.9 & 11.6 & 234.0 &  8.2 \\
1500 & 361.5 & 12.3 &  256.5 & 8.9 \\
1600 & 382.5 & 13.0 & 276.0 & 9.5 \\
\hline
\end{tabular}
\caption{Comparison between the maneuver and propellant mass fraction corresponding to a perigee-lowering strategy down to 80 km \cite{K06}, and our results to reenter from a quasi-circular
  orbit ($e=10^{-3}$) in 25 years.}
\label{Tab_Klink}
\end{table}
Moreover, if we recall Table \ref{Tab_Jan} from \cite{JetAl03}
discussed in Sec. \ref{sec:first_res}, we can observe that the
results shown in the fourth column of Table \ref{Tab_Klink} at
altitudes of $800-900\,$ km and $1200-1300\,$km are qualitatively in
agreement with the results shown in Table \ref{Tab_Jan} for the cases
of, respectively, a maximum available $\Delta v$ of 100 km/s or 200
km/s to reenter in 25 years.

Finally, we consider the results reported in \cite{W04} - Fig. 6,
where the propellant mass fraction is plotted as a function of the
required remaining lifetime for two circular orbits at altitudes
$H=800$ km and $H=1400$ km, respectively, and two ballistic
coefficients (20 and 200 kg/m$^2$, respectively, corresponding to
$A/m=0.05$ m$^2$/kg and $A/m=0.005$ m$^2$/kg), assuming a bit lower
exhaust velocity, $w_e=2550$ m/s ($I_{sp}=260$ s instead of
$I_{sp}=280$ s). From Table \ref{Tab_Klink} we find that our outcome
for a reentry in 25 years gives a required mass fraction of $2.7\%$ to
ensure a reentry from an initial quasi-circular orbit at $H=800$ km
(which grows to $2.9\%$ if we assume the same $w_e$ as in \cite{W04})
and a mass fraction of $8.2\%$ for a reentry from an initial orbit at
$H=1400$ km. These values are comparable or even better than the
results shown in Fig. 6 of \cite{W04}, supporting the advantages that
can be taken from this study.

\subsection{Results: high area-to-mass ratio}

To show some illustrative results in the case that an area
augmentation device is available on-board the spacecraft, we
consider two couples of initial orbits, whose orbital elements
together with the maximum $\Delta v$ available on-board are shown in
Table \ref{Tab:case_high}.
\begin{table}[htb!]
\centering
\begin{tabular}{cccccccc}
\hline
 & $a$ (km) & $e$ & $h_p$ (km) & $i$ ($^{\circ}$) & $\Omega$ ($^{\circ}$) & $\omega$ ($^\circ$) & $\Delta v$ (m/s) \\
\hline
Case 1a & 7900 & 0.001 & 1514 & 10 & 0 & 0 & 60 \\
Case 1b & 7900 & 0.001 & 1514 & 40.2 & 0 & 0 & 60 \\
\hline
Case 2a & 8170 & 0.01 & 1710 & 10 & 90 & 0 & 260 \\
Case 2b & 8170 & 0.01 & 1710 & 40 & 90 & 0 & 260 \\
\hline
\end{tabular}
\caption{Initial orbits selected as a reference case for
  $A/m=1\,$m$^2/$kg: semi-major axis, eccentricity, altitude of
  the perigee, inclination, right ascension of the ascending node,
  argument of perigee, maximum $\Delta v$ available on-board.}
\label{Tab:case_high}
\end{table}

The first two cases refer to a quasi-circular orbit at an altitude
$h=1522$ km and differ only in inclination. At this altitude, the
resonant inclination associated to $\dot\psi_1\simeq 0$ is
$i_{res}=39.7^{\circ}$. Thus, the inclination of orbit 1a is very far
from any resonance, while the inclination of orbit 1b is only
$0.5^{\circ}$ next to the resonant value. Moreover, we assume to have
a maximum $\Delta v$ of 60 m/s available on-board. The $\Delta v$
required at the apogee to lower the perigee altitude down to 120 km is
350 m/s, thus a direct reentry is not feasible. In the case of orbit
1a, the available $\Delta v$ can provide the maximum displacements
$\Delta a=-133$ km and $\Delta e=0.017$, corresponding to a perigee
altitude of the target orbit of $h_p=1250$ km, which is too high to
achieve a natural reentry within 25 years. For the initial orbit 1b,
SRP can, instead, be exploited. A maneuver of $\Delta v=27.6$ m/s
provides the following variations in the orbital elements:
$\Delta a=-60$ km, $\Delta e =0.0076$ and $\Delta
i=0.04^{\circ}$.
Then, at this target orbit the resonance due to SRP leads to a
variation in eccentricity of $\Delta e_{SRP}=0.091$, which ensures a
reentry within 25 years.

The two orbits labeled as case 2 have an altitude of $h=1792$ km and
differ, again, only in inclination. The resonant inclination
associated to $\dot\psi_1\simeq 0$ at this altitude is
$i_{res}=38.9^{\circ}$. For this case the maximum $\Delta v$ available
is significantly higher than the previous case: despite a value of 260
m/s is not realistic for most of the practical cases, we selected such
a high $\Delta v$ in order to identify a test case where reentry can
be achieved also without exploiting a resonance. The $\Delta v$
required at the apogee to lower the perigee altitude down to 120 km is
now 390 m/s, thus a direct reentry is not possible. Nevertheless, for
test case 2a it exists a solution to reenter in 25 years, which
consists in a tangential maneuver $\Delta v_t=-230$ km/s applied at
the apogee, corresponding to the variations $\Delta a=-533$ km and
$\Delta e=0.0659$. In the case of initial orbit 2b, we can exploit a
SRP resonance, even if the initial inclination is now $1^{\circ}$ next
to the resonant inclination. First, a single-burn maneuver of
$\Delta v=151$ m/s moves the spacecraft to a target orbit by means of
the following variations: $\Delta a=-344$ km, $\Delta e=0.042$ and
$\Delta i =0.207^{\circ}$.  Then, SRP drives a growth of eccentricity
up to $\Delta e_{SRP}=0.045$ and a reentry within 25 years is
achieved.

\section{Discussion and conclusions}
\label{sec:concl}

In this paper, we have presented a discussion on the possible
end-of-life disposal reentry strategies in LEO, exploiting one
impulsive maneuver and/or the effect of dynamical
perturbations. Taking advantage of the detailed LEO cartography
obtained by the authors in the framework of the H2020 ReDSHIFT
project, first of all we have presented a single-burn disposal
strategy based on the grid adopted for the cartography and shown in
Table \ref{grid}. Observing that the assumption to dispose the
spacecraft into a target orbit belonging to the grid can become too
severe, we have then presented a different but synergic disposal
strategy based on the physical explanation of the behavior revealed by
the cartography. In particular, we focus on the possibility of
exploiting dynamical resonances, mainly due to SRP, to facilitate a
reentry. We can conclude that, as long as the initial inclination of
the spacecraft is within or nearby a resonant corridor, the SRP
perturbation alone can be exploited to reenter if an area augmentation
device is available on-board, while in the case of low $A/m$ ratio we
can take advantage from the resonant behavior only in combination with
the effect of atmospheric drag. The clear benefit that can be achieved
by exploiting a resonance to reenter highlights the importance of
choosing a good initial condition from the very early phases of the
mission. The inclination, in particular, shall be selected, trying to
find a trade-off between mission objectives, operational constraints
and end-of-life opportunities. Given the cost of a plane change
maneuver and the narrow realm of the dynamical resonances, we
recommend a value close enough to a natural highway, towards which the
satellite could be moved at the end-of-life.

\begin{figure}[h!]
\begin{center}
	\includegraphics[width=0.7\textwidth]{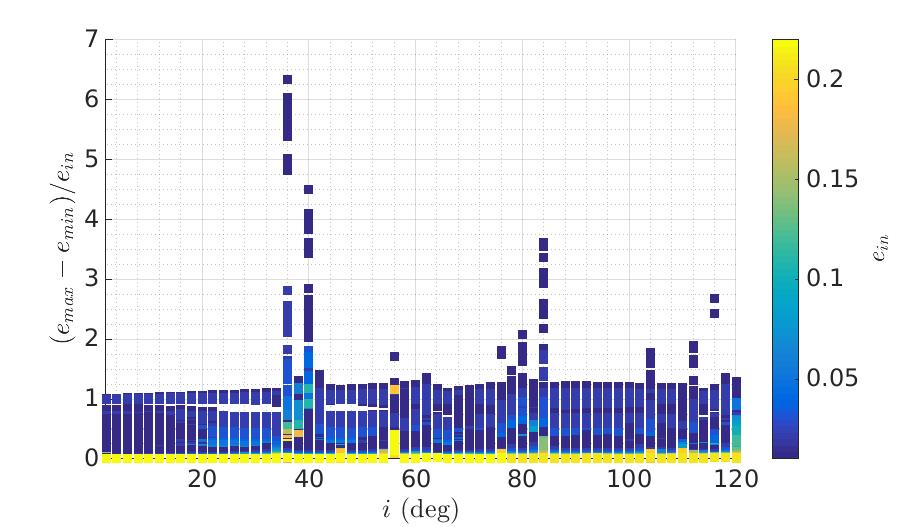}
\end{center}
    \caption{$(e_{max}-e_{min})/e_{in}$ for all the initial conditions such that
      $h_p>600$ km at the initial epoch 2020 and $A/m=0.012$
      m$^2$kg. The color bar shows the initial value of eccentricity.}
   \label{fig:derel_LEO}
\end{figure}

The analysis presented provides general important indications on how a
single-burn strategy shall be applied for mitigation purposes, both in
the perspective of a reentry and for a graveyard solution. As a matter
of fact, far from the resonances, the eccentricity can be considered
as constant, that is, the relative periodic variation is not
significant with respect to the initial value. This is shown in
Fig. \ref{fig:derel_LEO}, where we have represented the maximum
variation of the eccentricity relative to the initial value, i.e.,
$(e_{max}-e_{min})/e_{in}$ computed with FOP over 120 years, for all
values of initial semi-major axis such that the pericenter altitude is
higher than 600 km.
When the eccentricity does not experience dramatic changes, i.e.,
outside the resonance corridors, we expect that by changing the
initial epoch, the general behavior depicted, for example, in Fig.
\ref{fig:first_res} will not change. This also means that if a
spacecraft is left above the altitudes where the drag is effective and
outside the resonance corridors, it will stay there almost forever,
because its orbit can be considered stable. This information can be
positively used when the required $\Delta v-$budget to reenter is too
high, and thus a graveyard solution must be selected. In this case,
the best option consists in the closest circular orbit above an
altitude of 2000 km, outside the resonance corridors. Note also that,
in this case, the maneuver should be applied considering the spatial
density of the target region, or the criticality of the corresponding
shells (see \cite{R15} and \cite{B17}). In particular it is worth
stressing that, even if the residual propellant on-board would not
allow for a proper disposal maneuver (either towards a re-entry
solution or towards a super-LEO graveyard zone), the information on
the criticality of the surrounding altitude shells might suggest a
small maneuver to allow the positioning in a low-criticality nearby
shell, thus minimizing the long-term environmental impact of the
abandoned spacecraft.

Future directions include the possibility of targeting the conditions
described here by means of two or more maneuvers and evaluating the
collision risk experienced by specific reentry trajectories.

\section{Acknowledgements}
This work is funded through the European Commission Horizon 2020,
Framework Programme for Research and Innovation (2014-2020), under the
ReDSHIFT project (grant agreement n$^{\circ}$ 687500).

We are grateful to Camilla Colombo, Ioannis Gkolias, Kleomenis
Tsiganis, Despoina Skoulidou for the useful discussions on the
problem.

\end{document}